# Validating The Effectiveness of Electrospun Self-Healing Diels–Alder Interleaves to Mode-I fracture resistance by Comparing Simulation Outputs with Experimental Results

Constantinos Rouvalis[1], Vassilis Kostopoulos[1] and Spyridon Psarras[1]*

[1]Department of Mechanical Engineering and Aeronautics, University of Patras, GR-26500, Patras, Greece; corresponding author: spsarras@upatras.gr

## Abstract

The predictive capabilities of the finite element approach were assessed by comparing simulation outputs with experimental results, including load-displacement trends, damage initiation points, and delamination evolution. This comparison validated the effectiveness of the self-healing interleaves and highlighted the strengths and limitations of the adopted numerical framework. The simulations not only reproduced key damage characteristics but also provided a deeper understanding of failure mechanisms in the modified laminates. This modeling strategy contributes to the broader goal of developing high-fidelity virtual testing tools for complex, multifunctional composite structures used in aerospace and related industries.

## 1    Introduction

Fiber-reinforced polymer (FRP) composites have become indispensable across a range of industries due to their superior strength-to-weight ratio, corrosion resistance, and design flexibility. However, like all structural materials, FRPs are susceptible to damage, particularly in the form of matrix cracking and interlaminar delamination. Traditional repair techniques such as patching, resin injection, and surface coating offer temporary remedies but are often costly, labor-intensive, and incapable of addressing subsurface damage effectively [1].

To overcome these limitations, self-healing (SH) composites have emerged as a promising class of smart materials. These systems aim to autonomously or semi-autonomously repair internal damage, restoring functionality without external intervention [2, 3]. SH technologies are broadly categorized into autonomous systems—often relying on microcapsules or hollow fibers—and non-autonomous systems that respond to external stimuli such as heat or pressure. The latter approach, particularly those using thermally reversible chemistries like the Diels–Alder (DA) reaction, offers repeatable healing without the consumption of embedded agents [4-7].

Early SH research demonstrated the feasibility of embedding healing agents within hollow glass fibers (HGFs), mimicking biological responses to repair localized damage in polymers [4, 5, 7]. While effective, such extrinsic systems have limited repeatability and can compromise mechanical integrity if overused [8-10]. A more recent approach leverages the DA mechanism in bis-maleimide (BMI) polymers, which allows healing to be triggered by thermal activation and relies on reversible covalent bonds. This intrinsic healing method has shown excellent potential for aerospace composites, particularly in carbon fiber-reinforced systems [11-13].

To deliver these SH agents effectively within CFRP laminates, the Solution Electrospinning Process (SEP) has gained prominence. SEP enables the uniform deposition of nanofiber-based interleaves containing unreacted SH agents directly between prepreg plies prior to curing. This results in tailored interlaminar regions that maintain in-plane mechanical properties while introducing self-healing functionality [12]. Recent studies have further enhanced these systems using nanofillers such as multi-walled carbon nanotubes (MWCNTs) and graphene nanoplatelets (GNPs), which improve both mechanical and healing performance [14-16].

This study builds upon prior experimental work [17] by simulating the tensile behavior of CFRPs modified with DA-based SHAs applied via SEP. To capture the complex multi-mode failure mechanisms observed, including matrix cracking, fiber rupture, and interlaminar delamination, a finite element (FE) framework was developed using Abaqus/Explicit. The intralaminar damage response was modeled using Hashin's failure criteria, which enables the differentiation between fiber and matrix failure modes under tension and compression. Interlaminar delamination was represented using surface-based cohesive contact, a widely adopted approach that eliminates the need for predefined cohesive elements while maintaining numerical stability and physical accuracy [18-20].

The cohesive contact behavior was governed by bilinear traction-separation laws, calibrated against experimental Mode I delamination data. Recent studies highlight the effectiveness of such approaches in accurately reproducing delamination onset and propagation, especially when mesh sensitivity and cohesive strength parameters are carefully tuned [21, 22]. Moreover, the mesh architecture was optimized through convergence studies to resolve the high stress gradients near the open hole without incurring prohibitive computational cost. This modeling strategy has been successfully employed in similar studies to predict progressive damage in complex composite structures under both static and fatigue loading [19, 23, 24].

By comparing simulation results against experimental data—including load-displacement curves, failure modes, and delamination patterns—this study assesses the predictive capability of the adopted FE methodology. In doing so, it not only validates the performance of novel self-healing interleaves but also contributes to ongoing efforts to improve the fidelity of virtual testing methods for multifunctional aerospace-grade composites the numerical models that were developed – based on the Finite Element Method (FEM) with numerical Cohesive Zone Models (CZM) – are presented and their results are evaluated. Two dimensional (2D) models aim to approach the response of the examined materials under mode I fracture loading. The performed analyses were dynamic, with implicit time integration in the Standard module of Dassault Systèmes' Abaqus [25].

## 2   Materials and Methods

The geometric characteristics of the 2D models were used with the dimensions that acquired by the ASTM standard [14]. Moreover, two dimensional, 4-noded, plane strain finite elements (CPE4 in Abaqus) and two dimensional, 4-noded cohesive elements (COH2D4 in Abaqus) were utilized to discretize the geometry of the model. More precisely, the COH2D4 elements were placed in the five interlaminar regions that were modified with SHA, in order to account for the fracture response in these areas of the material. On the other hand, CPE4 finite elements were located in the tabs and the parts of the laminate (sublaminates) that were divided by cohesive layers.

The several parts of the total model (sublaminates, cohesive layers, tabs) were connected to each other by means of tie constraints. Additionally, the load and boundary conditions were applied in the middle points of the tabs. These points correspond to the central axes of the cylindrical pins that transmit the applied load to the specimen by connecting the grips of the testing machine with the tabs. At these points, the displacements along the x direction were constrained and a crosshead displacement load of 10 mm/min was applied in the y direction. Finally, the compliance of the testing machine was considered negligible.

## 2.1 Modeling the CFRP Sublaminates and Tabs with Finite Elements

As mentioned in the previous paragraph, the sublaminates of the DCB specimen and the tabs were discretized with 2D, 4-noded, plane strain finite elements (CPE4 in Abaqus). In Figure 2, the DCB specimen has been separated in three parts (A, B and C) along the x direction, in order to facilitate the presentation of the finite element dimensions. In the y direction, the regions of the sublaminates are noted as D, as shown in Figure 3. With respect to this notation, the dimensions or number of CPE4 elements are summarized in Table 1 for every region of the DCB specimen. As shown in this table, a finer mesh was created in region B, because it contains the edge of the precrack (Figure 3) and the 75 mm where the crack propagates during the experiment, until it reaches a total delamination crack length of 100 mm (including the 25 mm long precrack). As a result, this is the main region where the delamination phenomenon takes place and it is desirable to approach the deformation of the sublaminates with sufficient accuracy. Simultaneously, the creation of coarser meshes in regions A and C reduces the total computational cost of the numerical models.

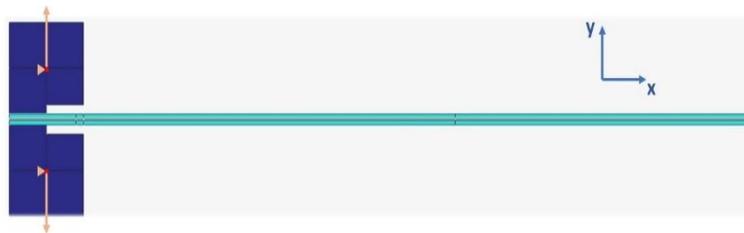

Figure 1: Applied load and boundary conditions on the tabs of the numerical models.

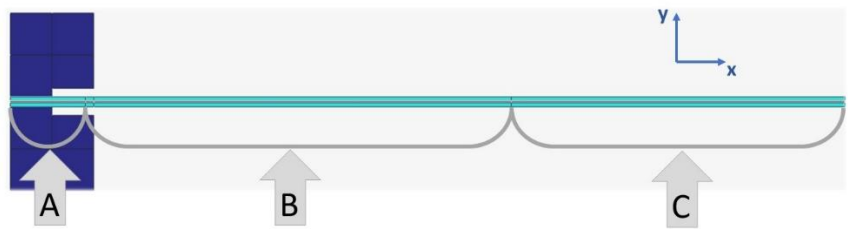

Figure 2: Separation of the DCB specimen geometry in regions A, B and C to describe the utilized element dimensions.

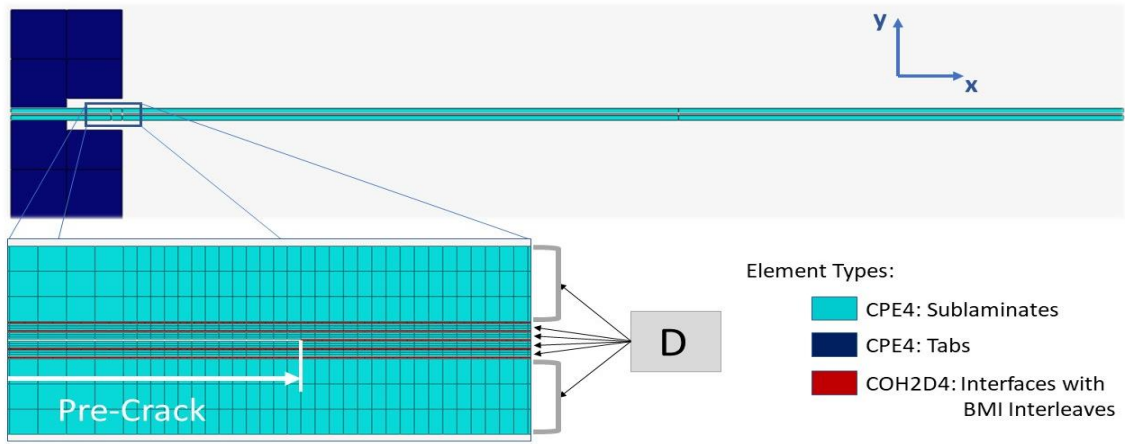

Figure 3: Regions D on a DCB specimen, along the direction of the thickness.

Table 1: Dimension or number of finite elements in every region of the model geometry for the DCB specimen.

| Region | Finite Element Dimension ($mm$) / No. of Finite Elements (elements) |
|--------|--------------------------------------------------------------------|
| A | 0,4 mm |
| B | 0,2 mm |
| C | 0,4 mm |
| D | 3 elements |

The UD CFRP plies that constitute the sublaminates of the specimen can be modeled as a transversely isotropic material. Therefore, five mechanical properties are required to define their mechanical behavior. Some of them were found in the material datasheet of the CFRP prepreg (SIGRAPREG C U150-0/NF-E340/38%) [26] and the rest were obtained from the literature [27]. The mechanical properties, the thickness and the density of the UD CFRP plies [28] are presented in Table 2. The ply thickness was useful to determine the positions of the five cohesive layers (that represent the SHA-modified interlaminar regions in the DCB specimens) along the y direction (Figure 3).

Table 2 Mechanical properties, density and thickness of a UD CFRP ply (SIGRAPREG C U150-0/NF-E340/38%) [26].

It has been experimentally proved [29] that the BMI – which plays the role of the SHA in the current study – has mechanical properties that are very similar to the mechanical properties of a thermoset epoxy resin. Consequently, it was assumed that the embedment of the BMI electrospun interleaves between the CFRP prepreg plies before curing and the infiltration of the BMI in the epoxy during the curing process did not affect the mechanical properties of the UD plies. Hence, the properties shown in Table 2. were assigned to the CPE4 elements of the sublaminates in the DCB specimen.

## 2.2    Quintilinear Cohesive Law for the SHA-Modified CFRP

When a large-scale fiber bridging mechanism is apparent during the fracture of a material, researchers in the literature have suggested the use of a cohesive law with the shape of Figure 4 [30-34]. In this type of cohesive law the cohesive stress has a linear relation with respect to separation up to maximum cohesive stress $\sigma_{max}^{C}$ without any damage, i.e., the damage variable is zero, D = 0. Afterwards, by a linear decreasing of the cohesive stress before maximum bridging stress $\sigma_{max}^{b}$, the damage grows at cohesive elements. The shaded area under this region (unhatched area) equals to the energy release rate for initiation of delamination. Following the linear part, the cohesive stress has a nonlinear behavior with respect to the separation (bridging law). Finally, at the end of the cohesive zone $\delta_f^*$, the cohesive stress vanishes, so the damage variable is equal to 1, D = 1. The hatched area under this region equals to the energy release rate that is attributed to fiber bridging.

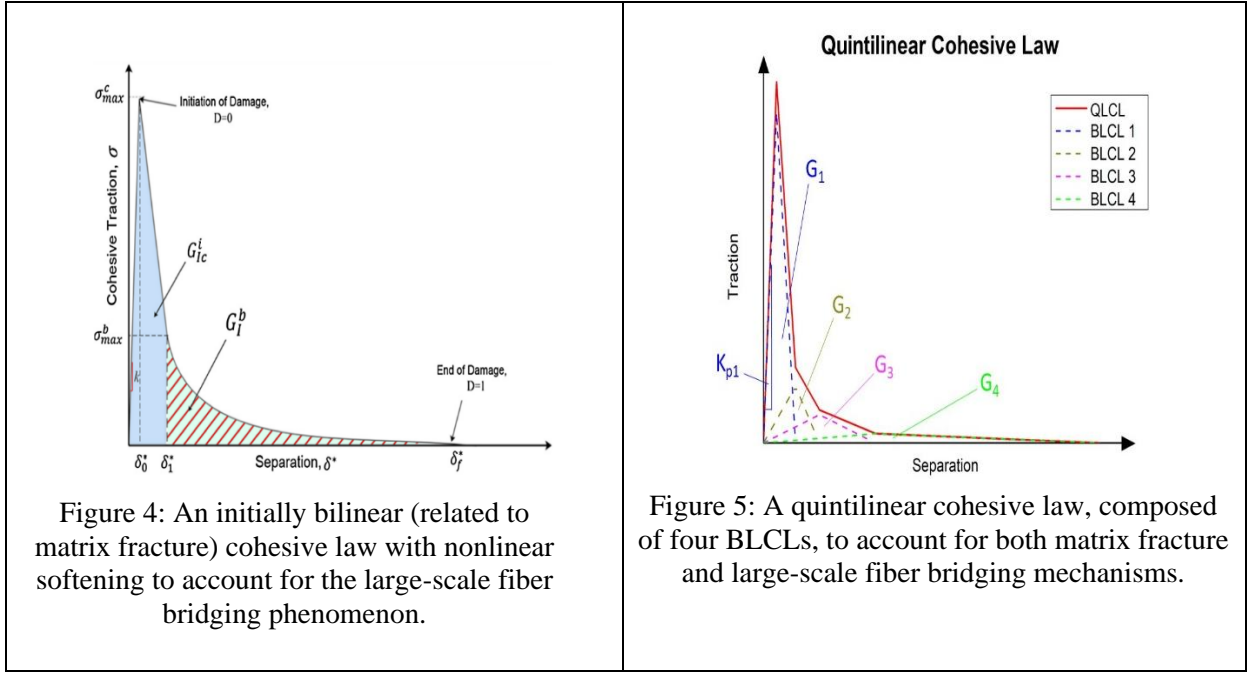

| Figure 4: An initially bilinear (related to matrix fracture) cohesive law with nonlinear softening to account for the large-scale fiber bridging phenomenon. | Figure 5: A quintilinear cohesive law, composed of four BLCLs, to account for both matrix fracture and large-scale fiber bridging mechanisms. |
|---|---|

Other research studies have proposed the use of multilinear cohesive laws to account for various types of fracture mechanisms with either small or large DPZ [23, 24, 35, 36]. In this type of cohesive laws, the parameters of the several linear parts – which constitute the multilinear law – are appropriately adjusted, so that the total shape of the law is suitable to describe the fracture mechanisms that are active during the evolution of the crack. A combination of the concepts for multilinear laws and the cohesive law that was presented in Figure 4 leads to the selection of a Quintilinear Cohesive Law (QLCL) to be utilized in the present study for the numerical analysis of the fracture response that was exhibited by the SHA-modified CFRP.

A QLCL is presented in Figure 5. Similarly to the superposition of two BLCL, the QLCL occurs by the superposition of four BLCL, while the maximum traction of the BLCL $(i + 1)$ is chosen to arise at the maximum separation of the BLCL $(i)$ and $i = 1, 2, 3$. Specifically, the response of the total QLCL emerges by superposing four layers of cohesive elements in every interlaminar region that is modeled with CZM and every cohesive layer is governed by a BLCL. This approach is encouraged by the fact that the representative R-curves of the SHA-modified CFRP (with or without nanofillers) can be approached by quintilinear curves, as shown in Figure 7, Figure 6 and Figure 8. Based on this approximation, the energy release rate $G_1$ is attributed to the brittle matrix fracture, it is equal to the fracture toughness of the material for crack initiation $G_{Ii}$ and is set as the fracture toughness of the BLCL 1 (Figure 5). The parts $G_2$, $G_3$ and $G_4$ (noted on the R-curves in Figure 7, Figure 6 and Figure 8) of the total fracture toughness $G_{IC}$ are related to the fiber-bridging phenomenon and are set as the fracture toughness of the BLCL 2, BLCL 3 and BLCL 4 (Figure 5), respectively. In total, the following equations hold:

$$G_{IC} = G_1 + G_2 + G_3 + G_4 \qquad \{2.1\}$$

$$G_{Ii} = G_1 \qquad \{2.2\}$$

$$G_{Ib} = G_2 + G_3 + G_4 \qquad \{2.3\}$$

,where $G_{Ib}$ is the fracture toughness that is attributed to the fiber-bridging fracture mechanism. The total approximation of the apparent fracture mechanisms with a QLCL that consists of four BLCLs is summarized in Figure 10.

The trial-and-error method was implemented and several numerical analyses were run to determine suitable parameters for the BLCLs 1, 2, 3 and 4 for the three material types which contain SHA, so that the criteria can be met. This process led to the selection of the parameters that are presented in Table 4, Table 5 and the force – displacement responses of the models with these parameters are drawn in Figure 11, Figure 12 and Figure 13. Moreover, an opening DCB specimen and its respective numerical model are depicted in Figure 9, where the correspondence is obvious between a layer of cohesive elements (which accounts for the fiber bridging mechanism) and the bridging fibers (which appear during the experiment). A color scale appears in this figure, which represents the values of the damage variable D – which is saved as SDEG in Abaqus [1] – of the BLCL 4 on the layer of cohesive elements. The cohesive elements in the red region have a damage variable value D = 1, which means that their stiffness is equal to zero and they have collapsed, similarly to the bridging fibers of this region in the experiment.

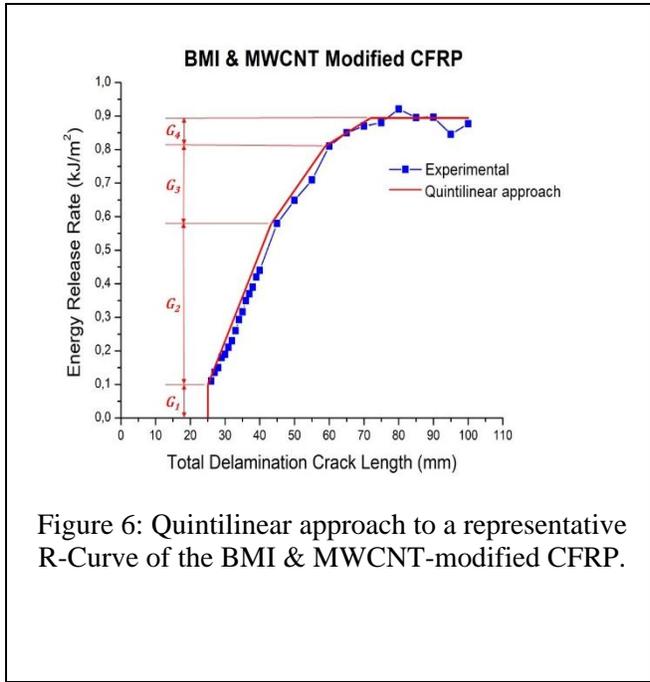

Figure 6: Quintilinear approach to a representative R-Curve of the BMI & MWCNT-modified CFRP.

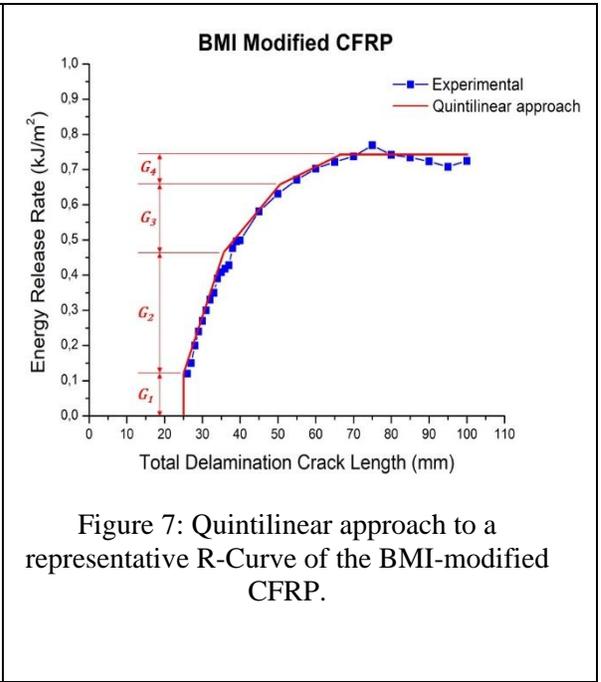

Figure 7: Quintilinear approach to a representative R-Curve of the BMI-modified CFRP.

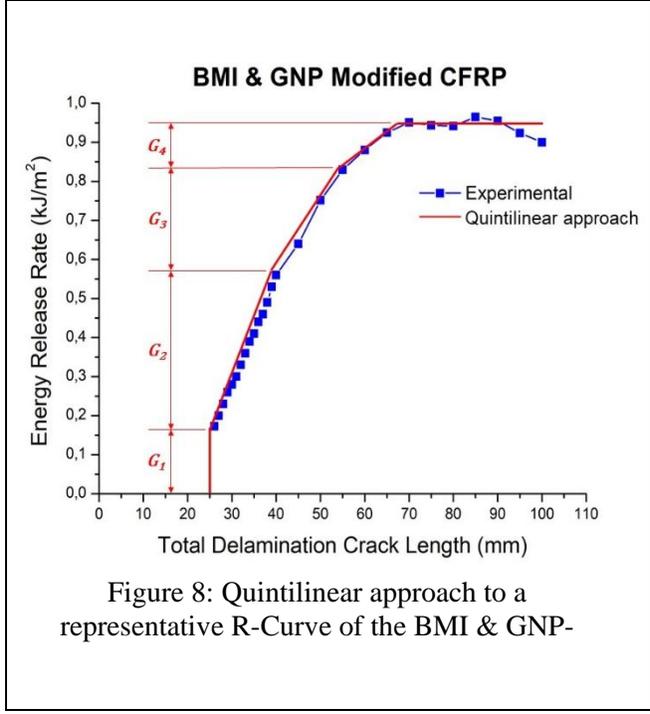

Figure 8: Quintilinear approach to a representative R-Curve of the BMI & GNP-

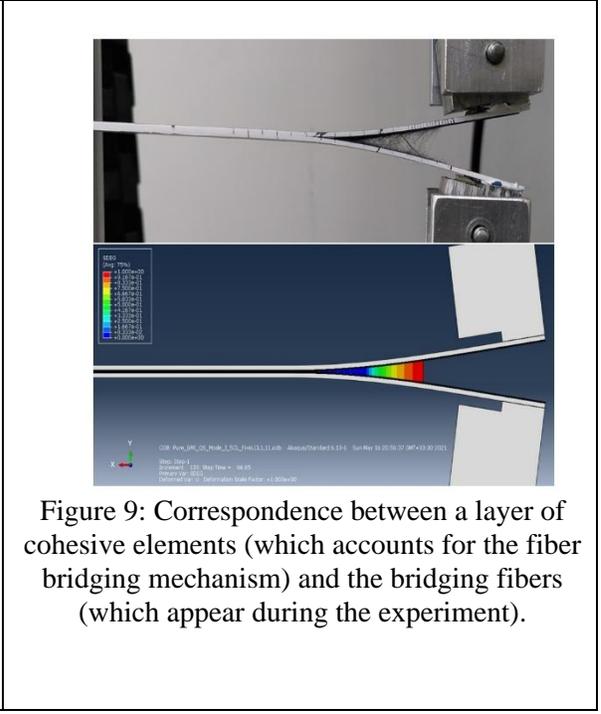

Figure 9: Correspondence between a layer of cohesive elements (which accounts for the fiber bridging mechanism) and the bridging fibers (which appear during the experiment).

The acceptance of acohesive law – during the execution of the trial-and-error method – as appropriate to describe thefracture response of each type of SHA-modified CFRP (with or without nanofillers) requires the satisfaction of the four following criteria:

a) The linear elastic stiffness in the force – displacement curve (i.e. the slope of the initial, linear part of the curve) of the numerical model should be as close as possibleto the maximum stiffness that has been exhibited among the specimens of the examined material type.

b) The peak force at the fracture of the matrix in the force – displacement curve of the numerical model should be as close as possible to the maximum peak force at matrix fracture that has been exhibited among the specimens of the examined material type.

c) The maximum force of the numerical model while the bridging phenomenon is apparent (after the fracture of the matrix) should be as close as possible to the maximum force that has been exhibited in this region of the force – displacement curve among the specimens of the examined material type.

d) The propagation part of the numerical force – displacement curve (from the displacement where the maximum load appears during fiber bridging, until the maximum displacement) should have a similar shape to the respective parts of the experimental curves of the examined material type and it should range between them.

The criteria (a), (b) and (c) for the SHA-modified CFRP emerge from the fact that the numerical models contained no material flaws and did not take into consideration any degradation or failure in the sublaminates of the specimen. Therefore, this ideal response was in closer agreement with the behavior of the specimens that demonstrated the highest values of stiffness and maximum load, i.e. they included the smallest amounts of defects.

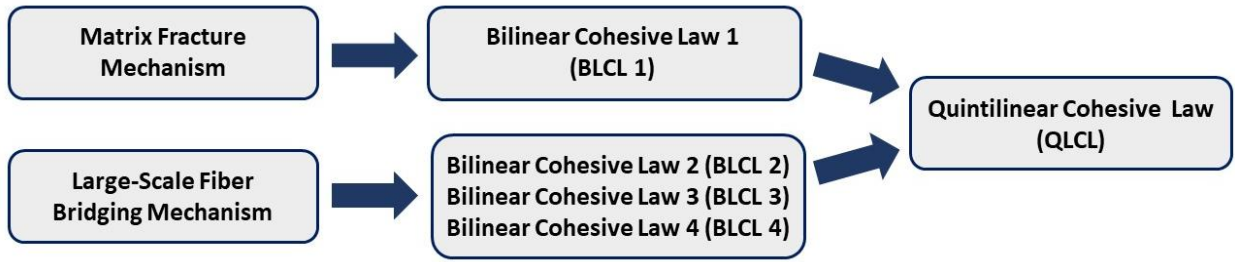

Figure 10: Correlation between the apparent fracture mechanisms and the BLCLs that compose the QLCL.

Table 3: Parameters of the four BLCLs that compose a QLCL for the pure BMI-modified CFRP.

| Quintilinear Cohesive Law Parameters – BMI-modified CFRP | | | |
|---|---|---|---|
| Bilinear Cohesive Law | Penalty Stiffness $K_p$ $(N/mm^3)$ | Fracture Toughness $G_I$ $(kJ/m^2)$ | Maximum Stress $\sigma_{Imax}$ (MPa) |
| BLCL 1 | $5 \times 10^3$ | 0,16 | 14 |
| BLCL 2 | $4,81 \times 10$ | 0,34 | 1,1 |
| BLCL 3 | $2,43 \times 10^{-1}$ | 0,13 | 0,15 |
| BLCL 4 | $2,88 \times 10^{-2}$ | 0,12 | 0,05 |

Table 4: Parameters of the four BLCLs that compose a QLCL for the BMI & MWCNT-modified CFRP.

| Quintilinear Cohesive Law Parameters – BMI & MWCNT-modified CFRP | | | |
|---|---|---|---|
| Bilinear Cohesive Law | Penalty Stiffness $K_p$ $(N/mm^3)$ | Fracture Toughness $G_I$ $(kJ/m^2)$ | Maximum Stress $\sigma_{Imax}$ (MPa) |
| BLCL 1 | $2,5 \times 10^3$ | 0,2 | 12 |
| BLCL 2 | $1,8 \times 10$ | 0,32 | 0,6 |
| BLCL 3 | $1,5 \times 10^{-1}$ | 0,2 | 0,16 |
| BLCL 4 | $2 \times 10^{-2}$ | 0,13 | 0,05 |

Table 5: Parameters of the four BLCLs that compose a QLCL for the BMI & GNP-modified CFRP.

| Quintilinear Cohesive Law Parameters – BMI & GNP-modified CFRP | | | |
|---|---|---|---|
| Bilinear Cohesive Law | Penalty Stiffness $K_p$ $(N/mm^3)$ | Fracture Toughness $G_I$ $(kJ/m^2)$ | Maximum Stress $\sigma_{Imax}$ (MPa) |
| BLCL 1 | $4,5 \times 10^3$ | 0,22 | 16 |
| BLCL 2 | $2,55 \times 10$ | 0,4 | 0,7 |
| BLCL 3 | $1,31 \times 10^{-1}$ | 0,15 | 0,15 |
| BLCL 4 | $2,5 \times 10^{-2}$ | 0,12 | 0,05 |

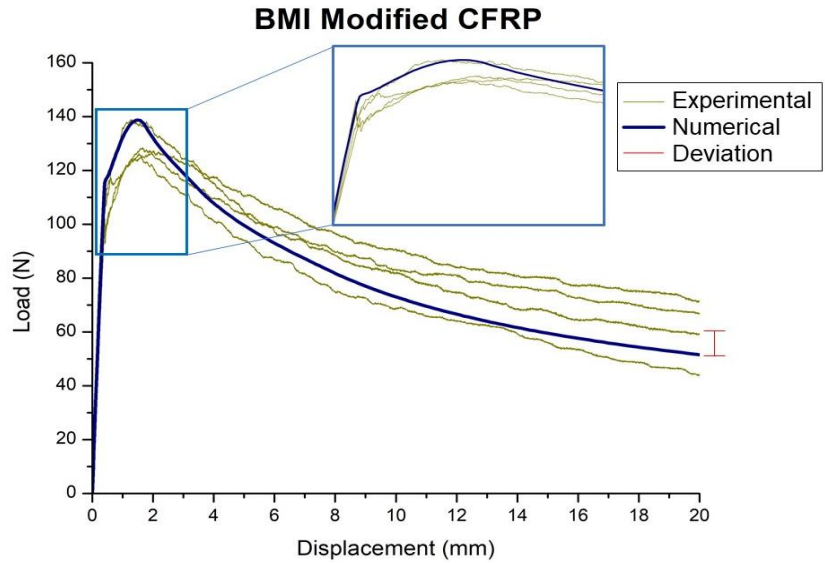

Figure 11: Numerical force – displacement response when a QLCL with suitable parameters is used to describe the response of the BMI-modified CFRP and comparison with the respective experimental curves.

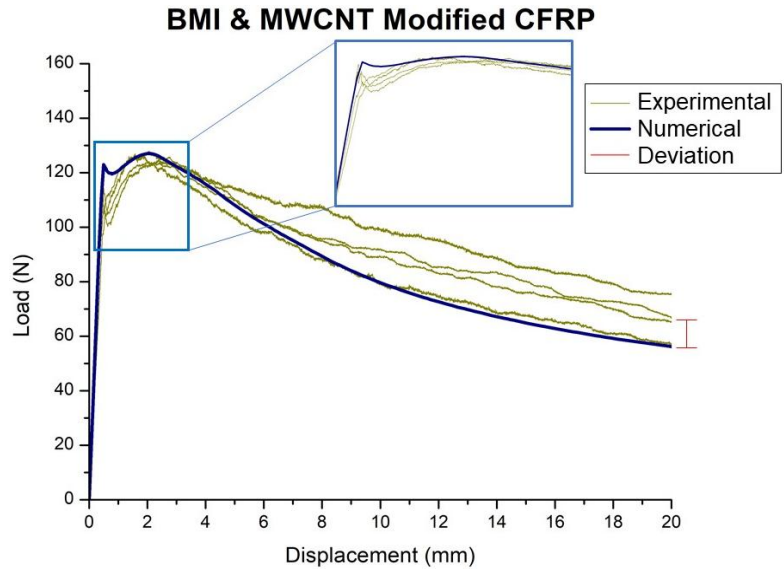

Figure 12: Numerical force – displacement response when a QLCL with suitable parameters is used to describe the response of the BMI & MWCNT-modified CFRP and comparison with the respective experimental curves.

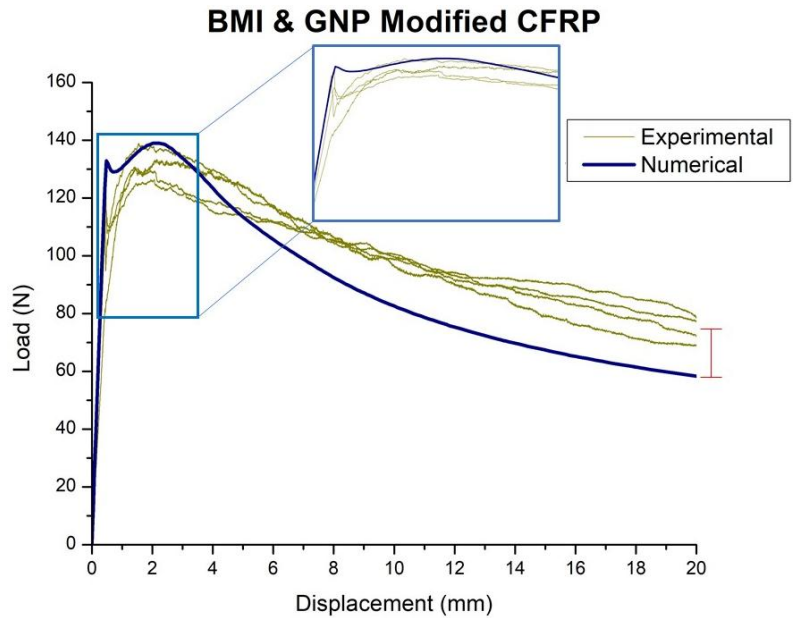

Figure 13: Numerical force – displacement response when a QLCL with suitable parameters is used to describe the response of the BMI & MWCNT-modified CFRP and comparison with the respective experimental curves.

If we observe Figure 11, Figure 12 and Figure 13, we can note that the initial, linear stiffnesses, the peak loads at matrix fracture and the maximum loads during bridging of the numerical curves are very close to the experimental ones. Hence, we can infer that the criteria (a), (b) and (c) are satisfied. However, only the shape of the propagation part of the numerical curve for the SHA-modified CFRP (without nanofillers) is similar to the respective parts of the experimental curves, but it tends to be closer to the lowest experimental curves. On the contrary, the propagation parts of the numerical curves for the SHA & MWCNT-modified and the SHA & GNP-modified CFRP have a different shape from the majority of their respective experimental curves. In addition, the values of the applied load in this part of the numerical curves are lower than the applied load on the respective part of the experimental curves. In consequence, we can conclude that the criterion (d) is not satisfied by the present approach.

## 2.3 Verification of Numerical Convergence

When a numerical method is implemented to analyze a given problem, it is important to verify the convergence of the obtained results. In other words, it is critical to make sure that the outcomes of the method are not dependent on the utilized mesh. If a Cohesive Zone Model is included in the numerical method, one of the main parameters that affect the numerical convergence is the length of the cohesive elements in the direction of crack propagation. In the published literature, meshes with cohesive elements which have a length of 0,1 mm (the used length in the previous analyses) in the region of crack propagation (region B of Figure 2) are considered as sufficiently fine to converge and analyze the majority of fracture mechanism with enough accuracy [24, 37, 38]. Hence, the investigation of the present paragraph is going to approve this fact for the examined models and simultaneously the convergence of the classical FEM will be verified.

Analyses with three different mesh densities were run, noted as Cases I, II and III in Table 6 and Table 7. For each case, the dimensions of finite and cohesive elements in the regions A, B, C and D of the models (Figure 2 and Figure 3) are included in Table 6 and Table 7. Case I uses an identical mesh to the one which was utilized in the previous paragraphs, while in Cases II and III the densities of the meshes are higher. In Figure 14, Figure 15 and Figure 16, the experimental and numerical force – displacement curves are included for all the material types and the examined mesh cases. In the legends of the figures, the numerical curves of cases I, II and III are denoted by their cohesive element length in the region B of the model (Figure 2), which is the cohesive element dimension that mainly affects the convergence of the CZM. The numerical curves of the three mesh cases for each type of material are coincident or too close to each other in these figures. Therefore, we can conclude that both the CZM and the classical FEM have sufficiently converged and the obtained results of the previous paragraphs (Case I) are mesh – independent.

Table 6: Dimension or number of finite elements in every region of the model geometry for the DCB specimen,in every case of mesh density, to verify the numerical convergence.

| Region | Finite Element Dimension ($mm$) / No. of Finite Elements (elements) | | |
|---|---|---|---|
| | Case I | Case II | Case III |
| A | 0,4 mm | 0,3 mm | 0,2 mm |
| B | 0,2 mm | 0,15 mm | 0,1 mm |
| C | 0,4 mm | 0,3 mm | 0,2 mm |
| D | 3 elements | 4 elements | 5 elements |

Table 7: Length of cohesive elements in every region of the model geometry for the DCB specimen, in every case of mesh density, to verify the numerical convergence.

| Region | Cohesive Element Length ($mm$) | | |
|---|---|---|---|
| | Case I | Case II | Case III |
| A | 0,2 | 0,15 | 0,1 |
| B | 0,1 | 0,075 | 0,05 |
| C | 0,4 | 0,3 | 0,2 |

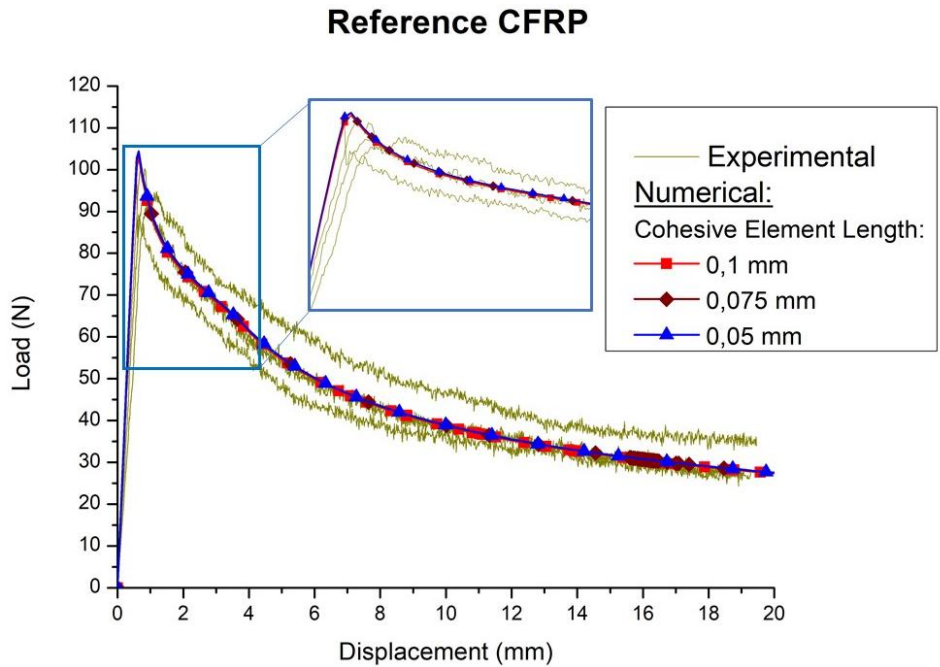

Figure 14: Experimental and numerical force – displacement curves with TLCL and various mesh densities for the reference CFRP.

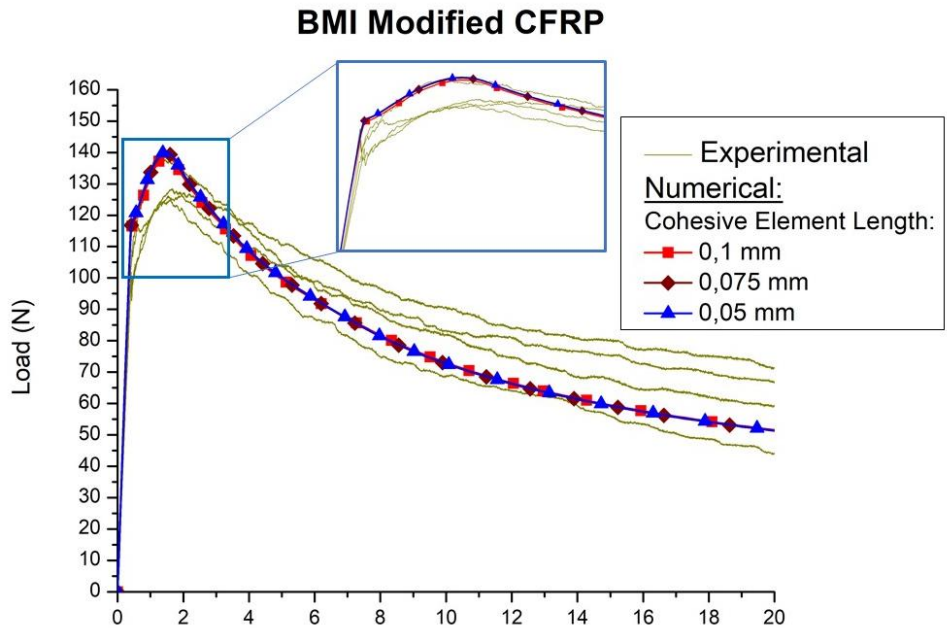

Figure 15: Experimental and numerical force – displacement curves with QLCL and various mesh densities for the BMI-modified CFRP.

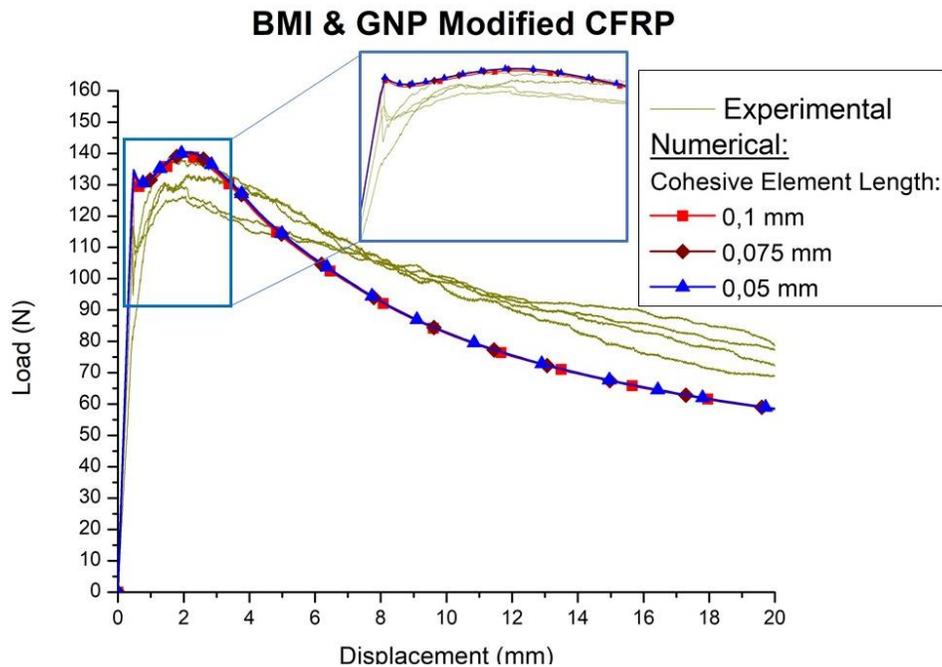

Figure 16: Experimental and numerical force – displacement curves with QLCL and various mesh densities for the BMI & MWCNT-modified CFRP.

## 3 Evaluation & Explanation of Deviation between the Numerical and Experimental Results

### 3.1 Evaluation of Numerical Results

Several parameters of the cohesive zone models have a direct relation with mechanical and fracture properties of the materials that were tested. As long as the developed models were based on experimental results, the evaluation of these parameters is going to facilitate the extraction of useful conclusions regarding the realistic, experimental properties of the materials. Specifically, the evaluated parameters are the penalty stiffness $K_{p1}$ of the BLCL 1, the fracture toughness $G_1$ of the BLCL 1 and the fracture toughness $G_{Ib}$. First, the penalty stiffness $K_{p1}$ explicitly affects the initial, linear stiffness of the force – displacement response that is exhibited by the model of the material. Besides, as mentioned in paragraph **Error! Reference source not found.**, the fracture toughness $G_1$ is equal to the fracture toughness for crack initiation $G_{Ii}$, which is related to the matrix fracture mechanism. Moreover, $G_{Ib}$ is the fracture toughness that is attributed to the fiber-bridging phenomenon and it occurs as the sum of $G_2$, $G_3$ and $G_4$ (equation {2.3}). The values of these parameters for the four tested types of materials are included in Table 8.

Table 8 Values of CZM parameters that are directly related to mechanical and fracture properties of the tested materials

| Type of CFRP | Initial Stiffness - $K_{p1}$ $(N/mm^3)$ | Matrix Fracture Mechanism - $G_1$ $(kJ/m^2)$ | Fiber-Bridging Fracture Mechanism - $G_{Ib}$ $(kJ/m^2)$ |
|---|---|---|---|
| Reference | 350 | 0,2 | — |
| BMI Modified | $5 \times 10^3$ | 0,16 | 0,59 |
| BMI & MWCNT Modified | $2,5 \times 10^3$ | 0,2 | 0,65 |
| BMI & GNP Modified | $4,5 \times 10^3$ | 0,22 | 0,67 |

First of all, the modeling parameters indicate that the modification of the reference CFRP with BMI interleaves (with or without nanofillers) enhances the initial, elastic stiffness of the material. The values of $K_{p1}$ are not a quantitative, but a qualitative indicator of the enhancement in the initial stiffness. More precisely, the modification with BMI (without nanofillers) and BMI & GNP had the most beneficial effect on the initial stiffness of the response, as their $K_{p1}$ values are close to each other and higher than the respective value of the BMI & MWCNT. Regarding the parameter $G_1$, it shows that the modification with electrospun interleaves containing BMI (without nanofillers) decreased the toughness of the matrix against delamination fracture in the CFRP. On the contrary, the incorporation of interleaves made out of BMI & GNP raised the matrix fracture toughness and the embedment of interleaves with BMI & MWCNT had a negligible impact on this property. Lastly, the parameter $G_{Ib}$ indicates that the inclusion of GNP in the BMI electrospun interleaves resulted in the fiber-bridging phenomenon that exhibited the highest fracture toughness, while the addition of no nanofillers in the BMI led to the fiber-bridging mechanism with the lowest fracture toughness. Totally, the chosen numerical parameters show that the BMI & GNP-modified CFRP displayed the most enhanced, aggregate mechanical and fracture response, which can be attributed to the augmented elastic stiffness and the increase of toughness in both the matrix fracture and the fiber-bridging mechanisms.

## 3.2    Explanation of Deviation between the Numerical and Experimental Results

As mentioned above, the criteria (c) and (d) for the similarity between the propagation parts of the numerical and the experimental force – displacement curves for all the material types were partially met. Consequently, despite the fact that the rest of the criteria  were satisfied, the reliability of the conclusions in the previous paragraph may be questioned and the numerical results cannot be characterized with certainty as absolutely realistic.

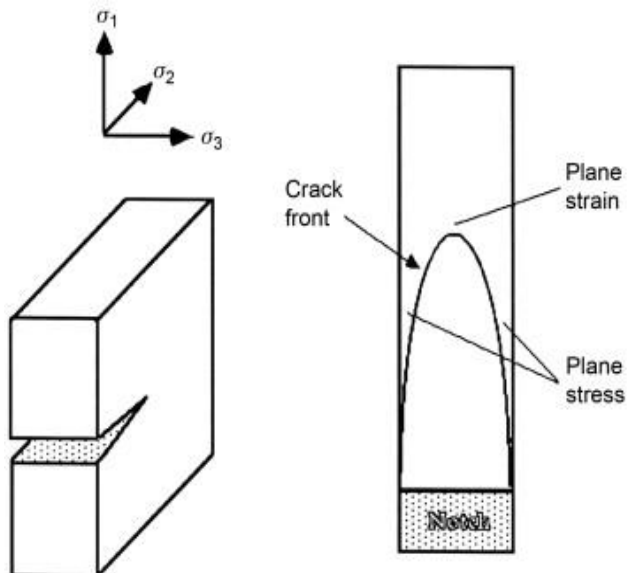

Figure 17 Shape of the crack front along the width of a DCB specimen

A reason which can be considered to cause this deviation between the experimental and the numerical results is the utilization of plane strain finite elements (CPE4) to approach the mechanical response of the sublaminates in the specimen (Figure 3). According to fracture mechanics, the front of a crack propagates inside a material in the U-like shape that is depicted in Figure 17 [39]. This shape occurs because the stress field in the vicinity of the crack approaches plane stress conditions close to the edges (free surfaces in the sides) of the cracked body and a plane strain state in the interior of the body. Particularly, the apparent fracture toughness of a material is higher under plane stress conditions and lower under plane strain, as described in Figure 18 [40]. Therefore, the crack propagates faster in the interior of the body and more slowly close to its edges, so it provokes the formation of the shape which is drawn in Figure 17 for the crack front. Based on these considerations, we can infer that the models mainly approached the mechanical and fracture response in the interior of the DCB specimens, close to their middle width, as long as plane strain elements were used. On the other hand, during the physical experiments, the DCB specimens contained plane stress regions close to their edges, which contributed to exhibit a larger resistance to the crack propagation and the cross-head displacement between the grips of the testing machine, in comparison to the resistance of the numerical models. For this reason, the majority of the experimental curves, Figure 11, Figure 12 and Figure 13, have a higher propagation part and require a higher level of applied force for the crack to propagate than the respective numerical curves, so that the criterion (c) and criterion (d) are not satisfied.

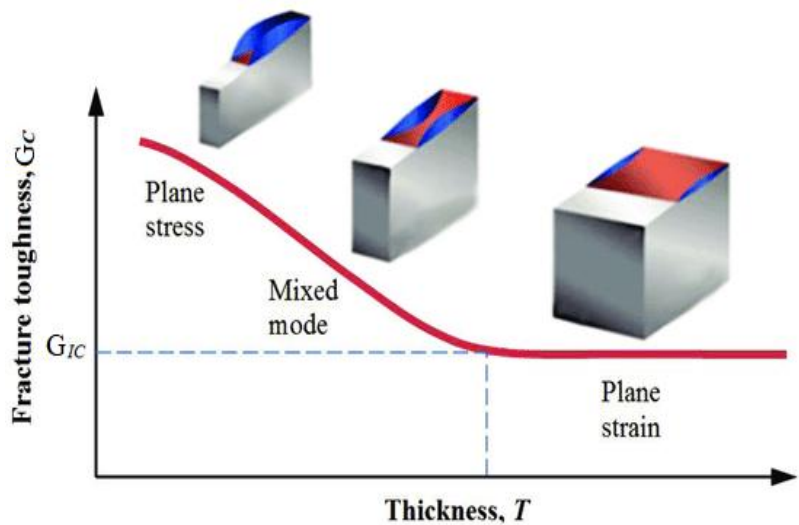

Figure 18: Dependence between the apparent fracture toughness of a body and its thickness.

In addition, one more observation that may be speculated as a cause of deviation between the numerical and experimental results is the emergence of damage in plies around the midplane of some DCB specimens, i.e. around the interlaminar region where the crack nominally propagates (Figure 19 ). This phenomenon was noticed only in the SHA-modified CFRP (with or without nanofillers) and only in a proportion of the total number of tested specimens for each material type (in about 2 out of the total 5 tested specimens for each material type). More precisely, in  Figure 19, the appearance of such damage is pointed with red arrows on four of the examined samples. The type of damage appears to be adjacent, delamination cracks in interlaminar regions above and under the midplane of the specimen. While the main, nominal delamination crack propagated in the midplane, one adjacent delamination crack evolved in every tested specimen which exhibited this phenomenon and lasted for a fraction of the total propagation length of the main crack (about 15 % to 20 % of the total, main delamination propagation length of 75 mm). The main delamination propagation lengths, where this phenomenon emerged, ranged between 30 mm and 60 mm. During the propagation of the main delamination, in front of its tip an adjacent crack initiated and grew until it reached a final length between 10 and 15 mm Afterwards, the adjacent crack remained stable and eventually closed when the main delamination propagated close enough, so from this point on the only open crack in the specimen was the main delamination, which kept on propagating until its final length of 100 mm (considering the precrack of 25 mm as a part of the final length).

As shown in Figure 19, the numerical models were unable to represent this phenomenon, as the delamination cracks appeared only in their midplane. Namely, the color scale of this figure describes the values of the damage variable (SDEG) in the cohesive elements of the model. As the delamination propagates towards the right side of the figure, the crack front is located in the region where the cohesive elements of BLCL 1 turn red (pointed with a blue arrow) and collapse (D = 1). The collapsed elements of the previous increment are deleted in every new increment of the analysis, so the blue cohesive elements on the left side of the red (collapsed) ones belong to BLCL 2, they account for the fiber-bridging phenomenon and are still intact. Simultaneously, the cohesive elements of BLCL 1 display no damage in the four interlaminar regions, which were modeled with CZM above and under the midplane. Thus, these regions remain unaffected by any type of adjacent damage, the numerical models did not represent the damage phenomenon which was described above and this may be an additional cause of deviation between the numerical and the experimental force – displacement curves for some of the tested specimens.

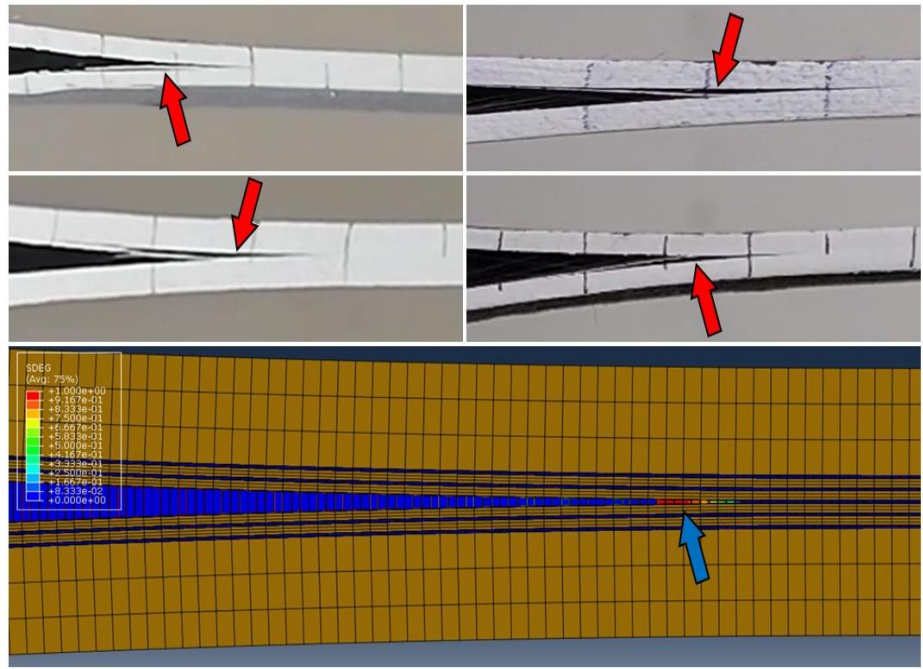

Figure 19: Top: Appearance of adjacent, delamination cracks (pointed with red arrows) in interlaminar regions above and under the midplane of the DCB specimen, during the experiments. Bottom: Appearance of delamination crack only in the midplane (pointed with blue arrow) of the DCB specimen in a numerical model, according to the SDEG variable.

## 4   Conclusions

Based on the conducted Mode I fracture experiments, FEM models were developed, including CZM to account for the fracture behavior of the interlaminar regions which contained electrospun interleaves. The models were only about the first fracture response of the specimens before healing, because after healing the SHA-modified CFRP (with or without nanofillers) showed insufficient recovery of fracture properties and a non-typical Mode I fracture behavior. Initially, bilinear cohesive laws were tested as constitutive relations of the cohesive elements and proved to be inappropriate to describe the apparent fracture response of the experiments. Subsequently, multilinear cohesive laws were chosen, in order to represent both the matrix fracture and fiber-bridging mechanisms that were noted during the experiments. Namely, a trilinear cohesive law was used for the response of the reference CFRP and Quintilinear cohesive laws were selected for the SHA-modified (with or without nanofillers) CFRP. The approximation with these models managed to meet almost all of the criteria that were set for the acceptance of numerical models as valid. Moreover, the chosen CZM parameters were in agreement with the experimental outcomes, regarding the superiority of the fracture properties of the SHA & GNP-modified CFRP over the rest of the examined material types. In addition, the mesh-independency of the obtained results was verified, by executing a convergence investigation, which confirmed that both FEM and CZM converged at a sufficient level. Nevertheless, the trueness of the models is questionable and they cannot be characterized as absolutely realistic. This fact is indicated by the deviations that appeared between the propagation parts of the numerical and the experimental force – displacement curves. Lastly, an attempt was made to explain the causes of these deviations, which were mainly attributed to the utilized plane strain approach of the stress field around the delamination crack and secondarily to the inability of the numerical models to represent adjacent delamination cracks, which emerged in some of the tested specimens.

## References

[1]     A. Baker, S. Dutton, and D. Kelly, *Composite Materials for Aircraft Structures*. AIAA Education Series, 2004.

[2]     D. G. Bekas, K. Tsirka, D. Baltzis, and A. S. Paipetis, "Self-healing materials: A review of advances in materials, evaluation, characterization and monitoring techniques," *Composites Part B: Engineering,* vol. 87, pp. 92-119, 2016.

[3]     D. Y. Wu, S. Meure, and D. Solomon, "Self-healing polymeric materials: A review of recent developments," *Progress in Polymer Science,* vol. 33, no. 5, pp. 479-522, 2008.

[4]     C. Dry, "Passive Tuneable Fibers and Matrices," *International Journal of Modern Physics B,* vol. 6, p. 2763, 1992.

[5]     C. Dry, "Procedures developed for self-repair of polymer matrix composite materials," *Composite Structures,* vol. 35, no. 3, pp. 263-269, 1996.

[6]     C. Dry, "Self-repairing, reinforced matrix materials," 2006.

[7]     C. Dry and N. Sottos, "Passive smart self-repair in polymer matrix composite materials," in *North American Conference on Smart Structures and Materials*, 1993: SPIE.

[8]     M. Hucker, I. Bond, A. Foreman, and J. Hudd, "Optimisation of Hollow Glass Fibres and their Composites," *Advanced Composites Letters,* vol. 8, no. 4, 1999.

[9]     M. J. Hucker, I. P. Bond, S. Haq, S. Bleay, and A. Foreman, "Influence of manufacturing parameters on the tensile strengths of hollow and solid glass fibres," *Journal of Materials Science,* vol. 37, no. 2, pp. 309-315, 2002.

[10]    G. Williams, R. Trask, and I. Bond, "A self-healing carbon fibre reinforced polymer for aerospace applications," *Composites Part A: Applied Science and Manufacturing,* vol. 38, no. 6, pp. 1525-1532, 2007.

[11]    V. Kostopoulos, Kotrotsos, A., Tsantzalis, S., Tsokanas, P., Christopoulos, A. C., Loutas, T. , "Toughening and healing of continuous fibre reinforced composites with bis-

maleimide based pre-pregs.," *Smart Materials and Structures,* vol. 084011, no. 25(8), 2016.

[12] V. Kostopoulos, Kotrotsos, A., Geitona, A., & Tsantzalis, S. , "Low velocity impact response and post impact assessment of CFRPs modified with Diels-Alder healing agent. ," *Composites Part A: Applied Science and Manufacturing,* no. 140, 106151, 2021.

[13] A. Kotrotsos, Tsokanas, P., Tsantzalis, S., Kostopoulos, V., "Healing of CFRPs by Diels–Alder polymers: Effects of SHA concentration and curing cycle," *Journal of Applied Polymer Science,* vol. 136(19), 47478, 2019.

[14] ASTM, "ASTM D7264 / D7264M-07. Standard Test Method for Flexural Properties of Polymer Matrix Composite Materials. ASTM International," 2007.

[15] P. Karapappas, Vavouliotis, A., Tsotra, P., Kostopoulos, V., & Paipetis, A. , "Enhanced fracture properties of carbon composites by adding multi-wall carbon nanotubes," *Journal of Composite Materials,* vol. 43(9), 977–985, 2009.

[16] C. Kostagiannakopoulou, Loutas, T., Sotiriadis, G., & Kostopoulos, V. , "Effects of graphene characteristics on CFRP interlaminar fracture toughness," *Engineering Fracture Mechanics,* vol. 245, 107584, 2021.

[17] A. Kotrotsos, C. Rouvalis, A. Geitona, and V. Kostopoulos, "Toughening and Healing of CFRPs by Electrospun Diels–Alder Based Polymers Modified with Carbon Nano-Fillers," *Journal of Composites Science,* vol. 5, no. 9, p. 242, 2021.

[18] P. P. Camanho, C. G. Davila, and D. R. Ambur, "Numerical Simulation of Delamination Growth in Composite Materials," *NASA Technical Report,* 2001.

[19] B. Cox and Q. Yang, "In quest of virtual tests for structural composites," *Science,* vol. 314, no. 5802, pp. 1102-1107, 2006.

[20] S. Sridharan, *Delamination Behaviour of Composites*. Woodhead Publishing, 2008.

[21] E. J. Barbero, *Finite Element Analysis of Composite Materials Using Abaqus™*. CRC Press, 2013.

[22] C. Rouvalis, "Comparison of Three Numerical Methods to Predict Delamination of Composites in Mode I Fracture Experiments: VCCT, CZM and XFEM," University of Patras, 2021.

[23] S. M. Jensen, M. J. Martos, E. Lindgaard, and B. L. V. Bak, "Inverse parameter identification of n-segmented multilinear cohesive laws using parametric finite element modeling," *Composite Structures,* vol. 225, p. 111074, 2019.

[24] S. Yin, Y. Gong, W. Li, L. Zhao, J. Zhang, and N. Hu, "A novel four-linear cohesive law for the delamination simulation in composite DCB laminates," *Composites Part B: Engineering,* vol. 180, p. 107526, 2020.

[25] "Dassault Systèmes, Abaqus 6.13 Documentation," ed, 2013.

[26] "SIGRAPREG C U150-0/NF-E340/38% datasheet. (2018). SGL Group."

[27] D. A. Türk, H. Einarsson, C. Lecomte, and M. Meboldt, "Design and manufacturing of high-performance prostheses with additive manufacturing and fiber-reinforced polymers," *Production Engineering,* vol. 12, no. 2, pp. 203-213, 2018.

[28] *SIGRAPREG C U150-0/NF-E340/38% datasheet* (SGL Group). 2018.

[29] T. Yuan, L. Zhang, T. Li, R. Tu, and H. A. Sodano, "3D Printing of a self-healing, high strength, and reprocessable thermoset," *Polymer Chemistry,* vol. 11, no. 40, pp. 6441-6452, 2020.

[30]     L. Sorensen, J. Botsis, T. Gmür, and L. Humbert, "Bridging tractions in mode I delamination: Measurements and simulations," *Composites Science and Technology,* vol. 68, no. 12, pp. 2350-2358, 2008.

[31]     E. Farmand-Ashtiani, D. Alanis, J. Cugnoni, and J. Botsis, "Delamination in cross-ply laminates: Identification of traction–separation relations and cohesive zone modeling," *Composites Science and Technology,* vol. 119, pp. 85-92, 2015.

[32]     G. Pappas and J. Botsis, "Intralaminar fracture of unidirectional carbon/epoxy composite: experimental results and numerical analysis," *International Journal of Solids and Structures,* vol. 85-86, pp. 114-124, 2016.

[33]     G. Frossard, J. Cugnoni, T. Gmür, and J. Botsis, "An efficient method for fiber bridging traction identification based on the R-curve: Formulation and experimental validation," *Composite Structures,* vol. 175, pp. 135-144, 2017.

[34]     M. M. Shokrieh, M. Salamat-talab, and M. Heidari-Rarani, "Dependency of bridging traction of DCB composite specimen on interface fiber angle," *Theoretical and Applied Fracture Mechanics,* vol. 90, pp. 22-32, 2017.

[35]     R. Vodicka and V. Mantic, "An energy based formulation of a quasi-static interface damage model with a multilinear cohesive law," *Discrete & Continuous Dynamical Systems,* vol. 10, no. 6, pp. 1539-1561, 2017.

[36]     S. M. Jensen, M. J. Martos, B. L. V. Bak, and E. Lindgaard, "Formulation of a mixed-mode multilinear cohesive zone law in an interface finite element for modelling delamination with R-curve effects," *Composite Structures,* vol. 216, pp. 477-486, 2019.

[37]     A. Turon, C. G. Dávila, P. P. Camanho, and J. Costa, "An engineering solution for mesh size effects in the simulation of delamination using cohesive zone models," *Engineering Fracture Mechanics,* vol. 74, no. 10, pp. 1665-1682, 2007/07/01/ 2007.

[38]     M. Moslemi and M. Khoshravan, "Cohesive Zone Parameters Selection for Mode-I Prediction of Interfacial Delamination," *Strojniški vestnik - Journal of Mechanical Engineering,* vol. 61, no. 9, p. 10, 2015.

[39]     M. D. Hayes, D. B. Edwards, and A. R. Shah, "4 - Fractography Basics," in *Fractography in Failure Analysis of Polymers*, M. D. Hayes, D. B. Edwards, and A. R. Shah, Eds. Oxford: William Andrew Publishing, 2015, pp. 48-92.

[40]     Š. Hajdu, "The Investigation of the Stress State near the Crack Tip of Central Cracks through Numerical Analysis," *Procedia Engineering,* vol. 69, pp. 477-485, 2014.